\documentclass[sigconf]{acmart} 
\AtBeginDocument{%
  }

\setcopyright{acmlicensed}
\copyrightyear{2025}
\acmYear{2025}
\acmDOI{XXXXXXX.XXXXXXX}
\acmConference[CIKM '25]{The 33rd ACM International
Conference on Information and Knowledge Management}{November 10--14,
  2025}{COEX, Seoul, Korea}
\acmISBN{978-1-4503-XXXX-X/2018/06}

\begin{document}

\title[Enhancing Serendipity Recommendation System by Constructing
Dynamic User Knowledge Graphs with LLMs]{Enhancing  Serendipity Recommendation System by Constructing Dynamic User Knowledge Graphs with Large Language Models}

\author{Qian Yong}
\authornote{Corresponding author}
\authornote{Both authors contributed equally to this research.}

\email{sarahyongq@gmail.com}
\affiliation{%
  \institution{Shanghai Dewu Information Group Co., Ltd.}
  \city{Shanghai}
  \country{China}
}

\author{Yanhui Li}
\authornotemark[2]
\email{liyanhui@stu.xjtu.edu.cn}
\affiliation{%
  \institution{Xi'an Jiaotong University}
  \city{Xi'an}
  \state{Shaanxi}
  \country{China}
}

\author{Jialiang Shi}
\email{shijialiang@dewu.com}
\affiliation{%
  \institution{Shanghai Dewu Information Group Co., Ltd.}
  \city{Shanghai}
  \country{China}
}

\author{Yaguang Dou}
\email{douyaguang@dewu.com}
\affiliation{%
  \institution{Shanghai Dewu Information Group Co., Ltd.}
  \city{Shanghai}
  \country{China}
}

\author{Tian Qi}
\email{qitian@dewu.com}
\affiliation{%
  \institution{Shanghai Dewu Information Group Co., Ltd.}
  \city{Shanghai}
  \country{China}
}

\renewcommand{\shortauthors}{Qian Yong, Yanhui Li, Jialiang Shi, Yaguang Dou and Tian Qi}

\begin{abstract}
The feedback loop of "system recommendation → user feedback → system re-recommendation" in industrial recommendation systems results in highly similar and convergent recommendation results.  This loop reinforces homogeneous content, creates filter bubble effects, and diminishes user satisfaction.  Recently, large language models (LLMs) have demonstrated potential in serendipity recommendation, thanks to their extensive world knowledge and superior reasoning capabilities.  However, these models still face challenges in ensuring the rationality of the reasoning process, the usefulness of the reasoning results, and meeting the latency requirements of industrial recommendation systems (RSs). To address these challenges, we propose a method that leverages large language models to dynamically construct user knowledge graphs, thereby enhancing the serendipity of recommendation systems.  This method comprises a two-stage framework: (1) two-hop interest reasoning, where user static profiles and historical behaviors are utilized to dynamically construct user knowledge graphs via large language models.  Two-hop reasoning, which can enhance the quality and accuracy of LLM reasoning results, is then performed on the constructed graphs to identify users' potential interests; and (2) Near-line adaptation, a cost-effective approach to deploying the aforementioned models in industrial recommendation systems. We propose a u2i (user-to-item) retrieval model that also incorporates i2i (item-to-item) retrieval capabilities, the retrieved items not only exhibit strong relevance to users' newly emerged interests but also retain the high conversion rate of traditional u2i  retrieval. Under offline evaluation criteria, the distribution of user potential interest scores is as follows: 1\% for a score of 0, 3\% for a score of 1, and 96\% for a score of 2.  Our online experiments on the Dewu app, which has tens of millions of users, indicate that the method increased the exposure novelty rate by 4.62\%, the click novelty rate by 4.85\%, the average view duration per person by 0.15\%, unique visitor click-through rate by 0.07\%, and unique visitor interaction penetration by 0.30\%, thereby enhancing user experience.
\end{abstract}



% \received{23 May 2025}
% \received[revised]{12 March 2025}
% \received[accepted]{4 August 2025}

\maketitle

\section{Introduction}
Recommendation systems have proven to be effective in alleviating information overload and providing personalized recommendation services to users \cite{ref1,ref2}. Currently, most recommendation systems employ deep learning algorithms trained on user historical interaction data \cite{ref3,ref4,ref5}. However, due to the feedback loop phenomenon of "system recommendation → user feedback → system re-recommendation", a significant portion of the content that users are exposed to originates from the system's previous recommendations, and user behavior data, in turn, affects the recommendation algorithm \cite{ref6}. This ultimately leads to both system recommendations and user behaviors being confined to a small subset of homogeneous content, a phenomenon known as the filter bubble effect \cite{ref2,ref7,ref8,ref9}. The filter bubble effect can result in users lacking surprising experiences and feeling bored and dissatisfied, thereby reducing user retention and app revenue. Some studies have attempted to mitigate these issues through diversified recommendation \cite{ref101,ref102,ref103}. However, such methods do not break through the users' fixed interest circles and therefore cannot fundamentally break the filter bubble effect.

To shatter the filter bubble and enhance user experience, serendipity recommendation has been proposed\cite{ref1,ref10}. Serendipity recommendation should present users with items that are not only unexpected but also relevant, thereby attracting user clicks and fulfilling both serendipity and relevance \cite{ref12,ref13}. In recent years, there has been significant progress in serendipity recommendation systems \cite{ref1,ref13,ref14,ref15,ref16,ref17}. However, due to the scarcity of serendipity data, these systems often resort to using smaller models or augmenting data based on biased recommendation data \cite{ref13}, which may inadvertently reinforce the feedback loop and increase the difficulty of breaking the filter bubble phenomenon and identifying novel items.

Recently, LLMs have been widely applied in various fields such as natural language processing \cite{ref18,ref19}, knowledge graphs \cite{ref20,ref21}, and recommendation systems \cite{ref7,ref23,ref24,ref25,ref26,ref27}. LLMs possess extensive world knowledge and have demonstrated remarkable capabilities in understanding and reasoning. LLMs can introduce external knowledge and interventions into recommendation systems to disrupt the filter bubble phenomenon. Moreover, recommendation systems based on LLMs are able to achieve outstanding performance with limited training data \cite{ref22}. Recent studies have attempted to leverage LLMs for serendipity recommendation and have achieved some results \cite{ref2,ref23,ref24}, yet they still face certain challenges.

Firstly, LLMs struggle to accurately generate answers to complex questions through single-hop reasoning. Inferring users' potential interests is a complex task. Simple prompts are insufficient to constrain the reasoning process of LLMs, leading to results that may lack novelty, have inappropriate levels of reasoning granularity, or be unrelated to user interests, failing to meet the requirements for novelty recommendation. The multi-agent debate\cite{refdebate} approach employs multiple language model instances to engage in multiple rounds of proposal and debate regarding their respective answers and reasoning processes, ultimately reaching a consensus answer. This method enhances the factual validity of the generated content and reduces the occurrence of hallucinations.

Secondly, industrial recommendation systems demand real-time performance, typically with response times within 100ms. Novelty recommendations based on large language models tend to have high latency and are computationally expensive. Efficiently deploying LLM-based serendipity recommendations in industrial recommendation systems remains a key challenge. Alibaba has proposed a near-line solution, SERAL\cite{ref7}, which is a pipeline system. SERAL first utilizes large language models to generate users' multi-level profiles and then integrates SerenGPT into the industrial RSs pipeline. The profile generation process is a two-stage task with high training costs and potential for cumulative errors.

Finally, after inferring users' potential interests, an important challenge in recommendation systems is how to efficiently retrieve relevant candidate items. Currently, the industry mainly adopts two approaches: (1) Directly utilizing the user's new interest embedding for i2i (item-to-item) retrieval, this is the simplest and most straightforward method, where top-k target items are retrieved via Approximate Nearest Neighbor (ANN) \cite{refann} or Maximum Inner Product Search (MIPS) \cite{refmips} based on the highest similarity to the trigger embedding; and (2) Training a u2i retrieval dual-tower model, which is a common industry practice due to its low cost and strong online performance. However, i2i retrieval suffers from the limitation of not leveraging the recommendation system's behavioral data, often resulting in suboptimal performance, such as low ranking model scores, poor post-retrieval metrics (e.g., PVCTR), and reduced recommendation efficiency for novel interests. Thus, i2i retrieved items are highly relevant to new interests but exhibit low conversion rate. On the other hand, u2i retrieval faces challenges because user interactions with new interests have not yet occurred, meaning these interests are absent from training data. Consequently, the model can only capture historical click interests. Even if new interests are incorporated as features in the user tower of a dual-tower model, they cannot be effectively represented. As a result, items retrieved via u2i  demonstrate high conversion rate while exhibiting limited relevance to users' new potential interests.

Therefore, we propose a novel approach to enhance the serendipity of recommendation systems by leveraging large models to construct dynamic user knowledge graphs. First, in two-hop interest reasoning, we use user static profiles and historical behaviors as input nodes to build dynamic user graphs with large models, and obtain users' potential interests through two-hop reasoning. Next, in nearline adaptation, we design a nearline surprise channel to generate and cache user interests, thereby avoiding the efficiency issues associated with online services of large language models. Then, in the retrieve stage, we design a multi-task learning dual-tower retrieval model: based on the traditional dual-tower u2i BCE-Loss, we also employ an interest-aligned contrastive learning loss in the user tower. Our contributions can be summarized as follows:

\begin{itemize}
 \item We design a two-stage  system to enhance the ability of large language models to identify serendipity and the feasibility of deployment, including constructing dynamic user knowledge graphs and performing two-hop reasoning to generate potential interests, and using multi-agent debate to ensure the correctness of the reasoning process and the relevance and serendipity of the results; as well as near-line adaptation to address deployment challenges.
 \item We propose a novel u2i retrieval model with integrated i2i retrieval capabilities. Specifically, the dual-tower retrieval model employs a multi-task learning framework, where, in addition to the traditional BCE loss in the user tower of a standard u2i dual-tower architecture, we introduce an interest-aware contrastive learning loss to enhance representation alignment.
 \item The system has been fully deployed online, resulting in an increase of 4.62\% in exposure novelty rate, 4.85\% in click novelty rate, 0.15\% in average view duration per person, and 0.30\% in unique visitor interaction penetration.

\end{itemize}

\section{Related work}

\subsection{LLM-based Recommendations}
In recent years, large language models have been widely applied in recommendation systems, with applications divided into two major categories: (1) Using LLMs as recommenders \cite{ref28,ref55,ref56,ref57,ref58,ref59}. Early research primarily focused on using LLMs for zero-shot recommendations \cite{ref69}, which works well for simple tasks but performs poorly in complex tasks. Subsequent studies shifted towards fine-tuning LLMs to infuse them with recommendation knowledge \cite{ref28,ref58,ref59}. (2) Integrating LLMs as components of traditional recommenders \cite{ref7,ref25,ref26,ref60,ref61,ref62,ref63,ref64,ref65,ref66,ref67}, which  not only boosts recommendation performance but also avoids the online inference latency issue of LLMs \cite{ref7,ref25,ref26,ref27,ref70}. For example, GENRE \cite{ref25} designs prompts for personalized news generation, user profiling, and news summarization, thereby improving the performance of various recommendation models. KAR \cite{ref27} and RLMRec \cite{ref26} leverage the open-world knowledge embedded in LLMs to enrich user behavior modeling in recommendation systems.
\subsection{Serendipity Recommendations}
Due to the feedback loops phenomenon, traditional recommendation systems tend to recommend items similar to previous suggestions, leading to user fatigue. To address this issue, serendipity recommendation has been proposed. Serendipity recommendations should include two key elements: unexpectedness and relevance \cite{ref23,ref71,ref72}. Essentially, recommendations should surprise users while also capturing their interest.

Serendipity recommendation methods can be divided into three categories: pre-processing, in-processing, and post-processing. Pre-processing methods focus on constructing serendipity features \cite{ref73,ref74}. In-processing methods integrate serendipity factors during model training and learn serendipity representations \cite{ref72,ref75,ref15,ref66,ref79}. Post-processing methods generate recommendation lists oriented towards serendipity \cite{ref78,ref79}. Among these, in-processing methods are the most prevalent. However, due to the scarcity of serendipity data, these methods often resort to using smaller models or augmenting data based on biased recommendation data. This may inadvertently reinforce feedback loops, complicating efforts to break through filter bubbles.

Recently, LLMs have been applied to serendipity recommendation \cite{ref23,ref24,ref73}. For example, LLMs can be used in a zero-shot manner to evaluate the serendipity of items \cite{ref24}. SerenPrompt \cite{ref23} crafts diverse prompts to utilize LLMs for serendipity assessment. SERAL \cite{ref7} employs LLMs to generate user profiles and make serendipity recommendations, aligning preference through human annotation and supervised fine-tuning. While these methods highlight the potential of LLMs in serendipity recommendation, they also reveal some unresolved challenges, such as the gap between LLMs and humans in judging serendipity and the real-time requirements for online services in industrial recommendation systems \cite{ref80,ref81,ref73}.
\begin{figure*}[htbp]
\centering
\includegraphics[width=0.8\textwidth]{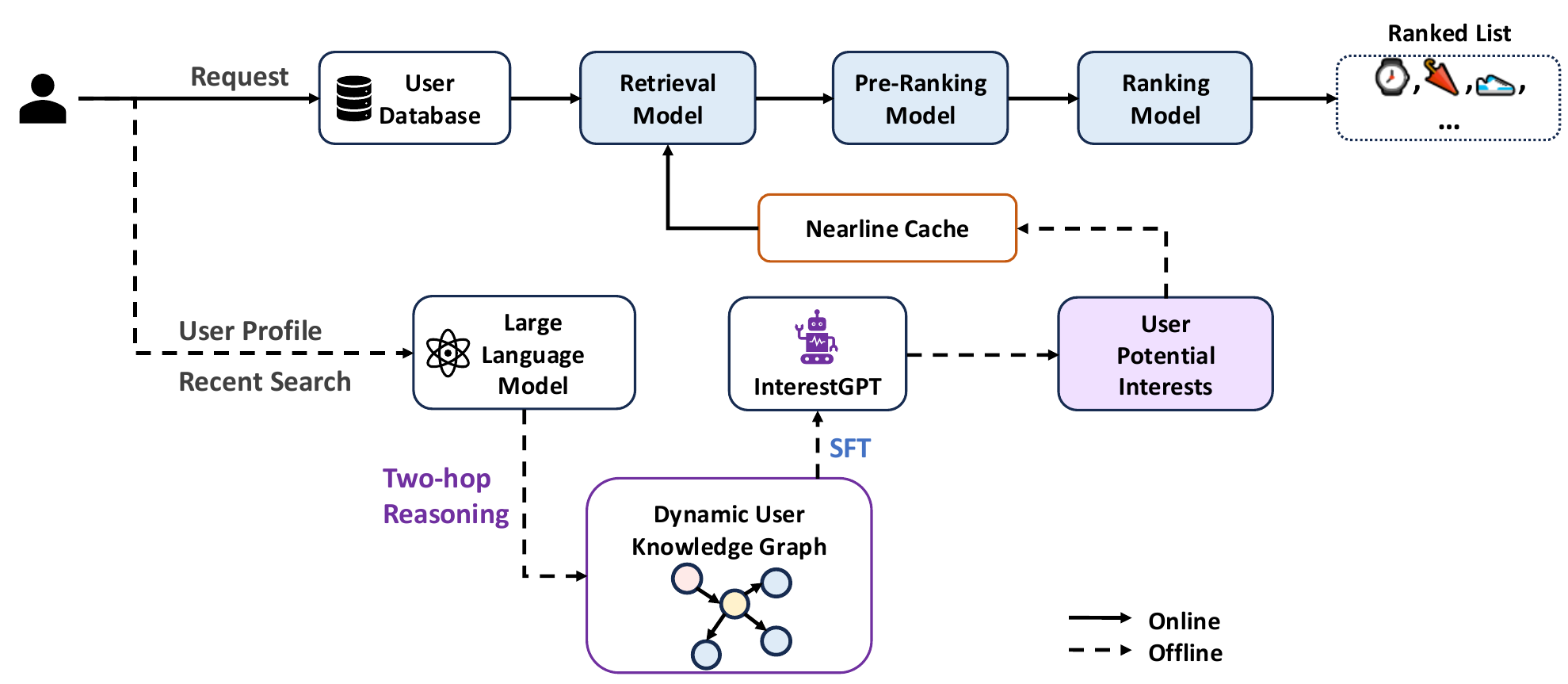}
\caption{An overview of our method for enhancing serendipity recommendation by constructing
dynamic user knowledge graphs with large language models.}
\label{overview}
\end{figure*}
\subsection{Two-hop Reasoning}
Multi-hop reasoning for knowledge graphs has been widely studied. This approach not only infers new knowledge but also generates interpretable paths, enhancing the credibility of the model \cite{ref36}. Most of the existing multi-hop reasoning models are based on the reinforcement learning (RL) framework. For instance,  DeepPath \cite{ref37} is the first work to formally propose and solve the task of multi-hop reasoning using RL. MINERVA \cite{ref38} proposed an end-to-end training framework, modeling path search as a sequential decision-making problem. M-Walk \cite{ref39} alleviated the sparse reward problem through off-policy learning, improving long-path reasoning stability. These methods rely on predefined knowledge graph structures but may encounter path disconnection issues when the knowledge graph is incomplete \cite{ref39,ref43}.

Recent studies have shown that LLMs possess excellent contextual reasoning capabilities \cite{ref46,ref47}. However, without explicit input of necessary information, LLMs struggle to perform multi-hop reasoning correctly, even if they know the correct answers for single-hop steps \cite{ref49}. Many researchers have implemented graph-like multi-hop reasoning on LLMs through prompt engineering. Chain-of-Thought (CoT) \cite{ref48} stimulates the model's implicit reasoning ability through step-by-step prompting but relies on explicit input information. Self-ask \cite{ref49} simulates the multi-hop querying process by explicitly decomposing sub-problems. Least-to-most \cite{ref50} achieves complex reasoning through problem decomposition, which has an intrinsic consistency with path search in knowledge graphs. KAR \cite{ref27} obtains open-world knowledge through factorized prompting.

\subsection{Multiagent Debate}
A wide range of research has explored how to achieve reasoning and factuality in language models. To enhance reasoning, some approaches have relied on prompting techniques, such as scratchpad \cite{ref82}, verification \cite{ref83}, chain-of-thought demonstration \cite{ref48,ref84,ref85}, intermediate self-reflection \cite{ref86}, and fine-tuning \cite{ref87,ref88,ref89}. To improve factuality, some methods depend on training techniques, such as RLHF \cite{ref90,ref91,ref92}, pruning truthful datasets \cite{ref93}, external knowledge retrieval \cite{ref94}.

Some studies have proposed an alternative approach to achieving reasoning and factuality in language models through multi-agent debate, which requires only black-box access to a language generator. Other works have explored majority voting among different models \cite{ref96,ref83,ref97,ref98}. Irving et al. \cite{ref99} also introduced a debate procedure to verify the accuracy and safety of powerful AI agents. By facilitating communication between different language models, these approaches enable more effective reasoning and factuality in language models. Multi-agent Debate \cite{refdebate} introduces a debate framework based on multi-agent collaboration. When faced with a problem, the system creates multiple parallel instances of large language model agents. Each agent initially generates its own response and reasoning process, reflecting different solution strategies or sources of information. After collecting the initial responses from all agents, the system initiates a multi-round debate mechanism. In this process, each agent critically evaluates the responses of the other, identifies any flaws in their reasoning, and refines its own response. This iterative cross-validation and self-correction process enhances the accuracy and reliability of the final answer.

\section{Task Formulation} 
Given an item set $I$ and a user set $U$, for the $i-$th user $u_i \in U$, the corresponding search behavior sequence is $Q_i=\{q_1,q_2,...,q_m\}$, where $q_j (j=1,2,...,m)$ represents the latest $j-$th query of user $u_i$. The goal of serendipity recommendation is to find items with serendipitous properties for user $u_i$. We divide this process into two steps. First, a large language model is used to generate potential interest keywords for user $u_i$ based on $Q_i$:

\begin{equation}
    P_i=\Pi_{\theta}(Q_i,S_i),
\end{equation}
where $P_i=\{interest_{i,1},interest_{i,2},...,interest_{i,n}\}$ is the set of potential interests for user $u_i$ generated by the large language model $\Pi_{\theta}$, $S_i$ denotes the user profile of $u_i$.

A text encoder is used to encode the user’s potential novel interests $interest_k \in P_i$ into user interest embeddings $z_{i,k}^u$.
%For $interest_k \in P_i$, we perform feature embedding on them:
\begin{equation}
   z_{i,k}^u= E(interest_k), \quad 1\leq k \leq n,
\end{equation}
where $E(\cdot)$ denotes the encoder.

During online retrieval, the potential interest embedding $z_{i,k}^u$ and user features $f_i^u$ are input into the retrieval model to retrieve $l$ items related to the $k-$th potential interest:
\begin{equation}
    \{item_{k,1}, item_{k,2}, \ldots, item_{k,l}\} = \text{Retrieve}(z_{i,k}^u, f_i^u).
\end{equation}

\section{Method}

\subsection{Overview}
The workflow of our method is depicted in Figure \ref{overview} and consists of two stages: offline generation of user latent interests and online serendipity retrieval. First, we dynamically construct a user knowledge graph based on user static profiles and historical behaviors using a large language model, and perform two-hop reasoning on the constructed graph to obtain user potential interests. This process yields InterestGPT through supervised fine-tuning. The reasoning results represent the user's potential interests, which are cached in nearline storage. Subsequently, during the online retrieval phase, we leverage the nearline-cached potential interests and employ a multi-task u2i retrieve model to retrieve items that are relevant to these potential interests.

\subsection{Dynamic User Knowledge Graphs-based Two-hop Reasoning }
\begin{figure}[htbp]
  \centering
  \includegraphics[width=\linewidth]{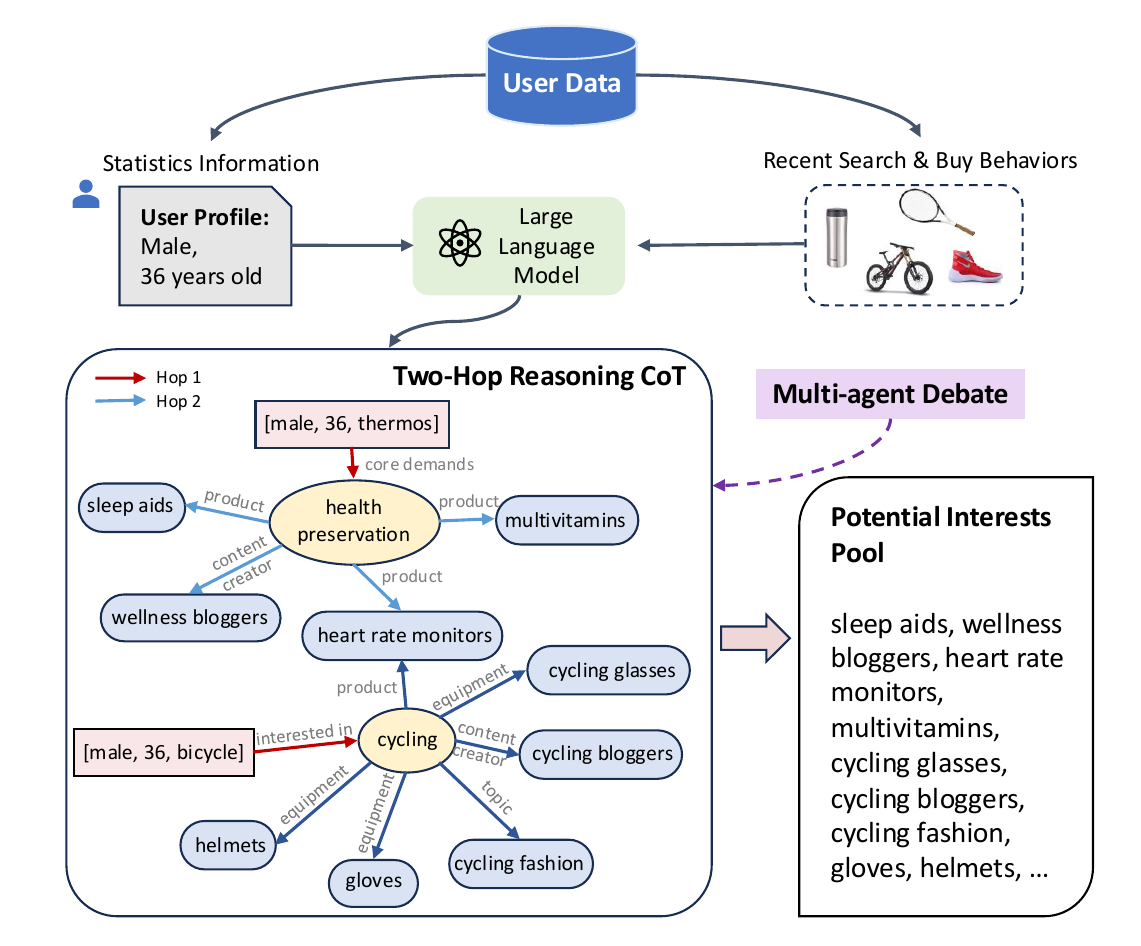}
  \caption{Leveraging LLMs to construct dynamic user knowledge graphs and performing two-hop reasoning.}
  \label{twohop}
\end{figure}

As shown in Figure \ref{twohop}, the user’s static profile (age and gender) and historical behaviors (queries in the past 30 days) are used as the initial input nodes. The large language model is used as a knowledge graph builder to dynamically construct nodes and relationships $\quad G=(V,E)$, where $V$ is the set of entities and $E$ is the set of relationships. Given two entities $v_1$ and $v_3$, the objective is to determine whether there exists a potential interest relationship between them through two-hop reasoning.

\textbf{Step 1:} Starting from the user’s static profile and query $v_1$, identify nodes $v_2$ that satisfy the hypernym relationship. Find all $v_2$ such that $(v_1,v_2) \in E$. $v_2$ represents the core demands and motivations of $v_1$.

\textbf{Step 2:} Starting from $v_2$, identify all nodes $v_3$ that satisfy the user’s core demands in a hyponym or co-hyponym relationship. Find all $v_3$ such that $(v_2,v_3) \in E$. To avoid irrelevant outputs and reduce hallucinations, $v_3$ is restricted to the scope of products, product categories, and topics.

\subsection{Multi-agent Debate}
Prompt engineering, which constructs dynamic user profiles and performs two-hop reasoning based on static user data and behaviors, may result in reasoning path errors and irrelevant potential interest outcomes. In this paper, we introduce a complementary method to enhance language responses. Specifically, multiple language model instances engage in a multi-round process of proposing and debating their individual responses and reasoning processes to reach a consensus final answer. We find that this approach significantly boosts the two-hop reasoning capability of the task. Additionally, our method enhances the factual accuracy of the generated content, mitigating the fallacious answers and hallucinations that contemporary models often produce.

\begin{figure}[htbp]
  \centering
  \includegraphics[width=\linewidth]{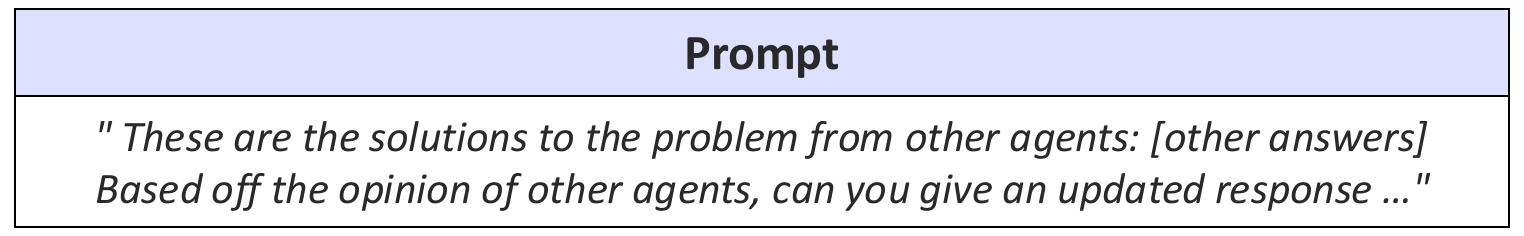}
  \caption{The prompt for inducing multi-agent debate.}
  \label{prompt}
\end{figure}

Concretely, we first prompt each agent to independently address the given problem or task. After each agent generates a response, we provide each agent with a consensus prompt, as depicted in Figure \ref{prompt}. Each agent is then instructed to refine their response based on the responses of the other agents. This consensus prompt can be iteratively reapplied, utilizing the updated responses from each agent in subsequent rounds. Our prompt is available on GitHub \footnote{Prompt for constructing dynamic user knowledge graph and SFT training data samples are available on GitHub: \url{https://github.com/DouYuoD/LLM-Two-hop-Reasoning-Recommendation} \label{link}}.

\subsection{Supervised Fine-tuning}
In order to reduce deployment costs, we first employ the high-capacity inference model DeepSeek-R1 \cite{deepseek} to construct the dynamic user knowledge graph and generate potential interests. Subsequently, we distill this information into a more compact model, QWQ-32B. The thought process and potential interests\footref{link} are then converted into a text-based supervised fine-tuning (SFT) dataset $D$, with each entry represented as a tuple $(x, y)$. Here, $y$ denotes the output, encapsulating the thought process and potential interests, while $x$ stands for the input prompt. Next, adhering to Eq. (\ref{eq:sft_loss}), we conduct supervised fine-tuning on QWQ-32B to derive InterestGPT, thereby enhancing its likelihood of generating the desired responses.
\begin{equation}
    \begin{aligned}
    \mathcal{L}_{\text{SFT}} &= -\mathbb{E}_{(x,y) \sim \mathcal{D}} \left( \log \pi_{\theta}(y|x) \right) 
    \\&= -\mathbb{E}_{(x,y) \sim \mathcal{D}} \left( \sum_{t=1}^{T} \log \pi_{\theta}(y_t | x, y_{1:t-1}) \right).
    \end{aligned}
    \label{eq:sft_loss}
\end{equation}

\subsection{Nearline Adaptation}
Large language models  face the issue of reasoning latency, which makes it difficult to meet the real-time requirements of recommendation system online services. Additionally, serendipity recommendation has a relatively weak dependence on users' real-time behavioral information. Its objective is to recommend novel yet appealing content to users, focusing more on long-term user experience rather than immediate demands. Over-reliance on real-time behavior may actually reduce the serendipity of recommendations.

Nearline computing is a compromise between real-time online processing and batch offline computation, achieving a balance between efficiency and timeliness. Therefore, we employ a nearline approach to infer users' potential interests, with the generated content being utilized in the online retrieval phase.

Specifically, we have introduced a potential interest  retrieval channel. First, based on users' historical search behaviors and basic information, we use LLMs to infer their potential interest keywords. The results are then stored in a nearline cache for use in online retrieval. This cache retains these potential interest keywords within a time window of $T$ days, with $T$ typically set to 7 days. Once the potential interest keywords expire, they will be recalculated. Moreover, when users engage in new search activities, the LLM will be invoked again to update the potential interest keywords.

\subsection{Retrieval Model}
\begin{figure}[htbp]
  \centering
  \includegraphics[width=\linewidth]{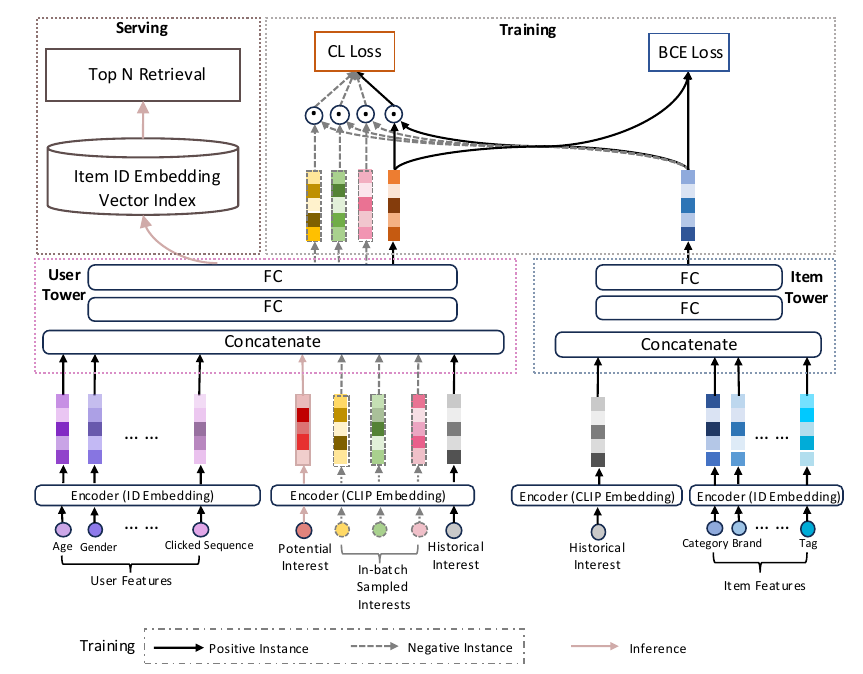}
  \caption{Retrieval Model.}
  \Description{Retrieval model.}
  \label{retrievemodel}
\end{figure}

To combine the strengths of both i2i and u2i retrieval, we designed a u2i retrieval model that also possesses i2i retrieval capabilities. Specifically, the dual-tower retrieve model is formulated as a multi-task objective. On the basis of the traditional dual-tower u2i BCE-Loss, we introduced a contrastive learning loss based on interest alignment in the user tower. By maximizing the similarity between user embeddings and item embeddings under the same interests, while minimizing the similarity between user embeddings and item embeddings under different interests, thereby enabling the generation of highly relevant user-embeddings based on the user’s new interests during the inference stage. These embeddings are used for u2i retrieval. The retrieved item set not only maintains high relevance to the potential interests but also retains the high conversion rate of u2i retrieval.

\begin{figure*}[htbp]
\centering
\includegraphics[width=0.9\textwidth]{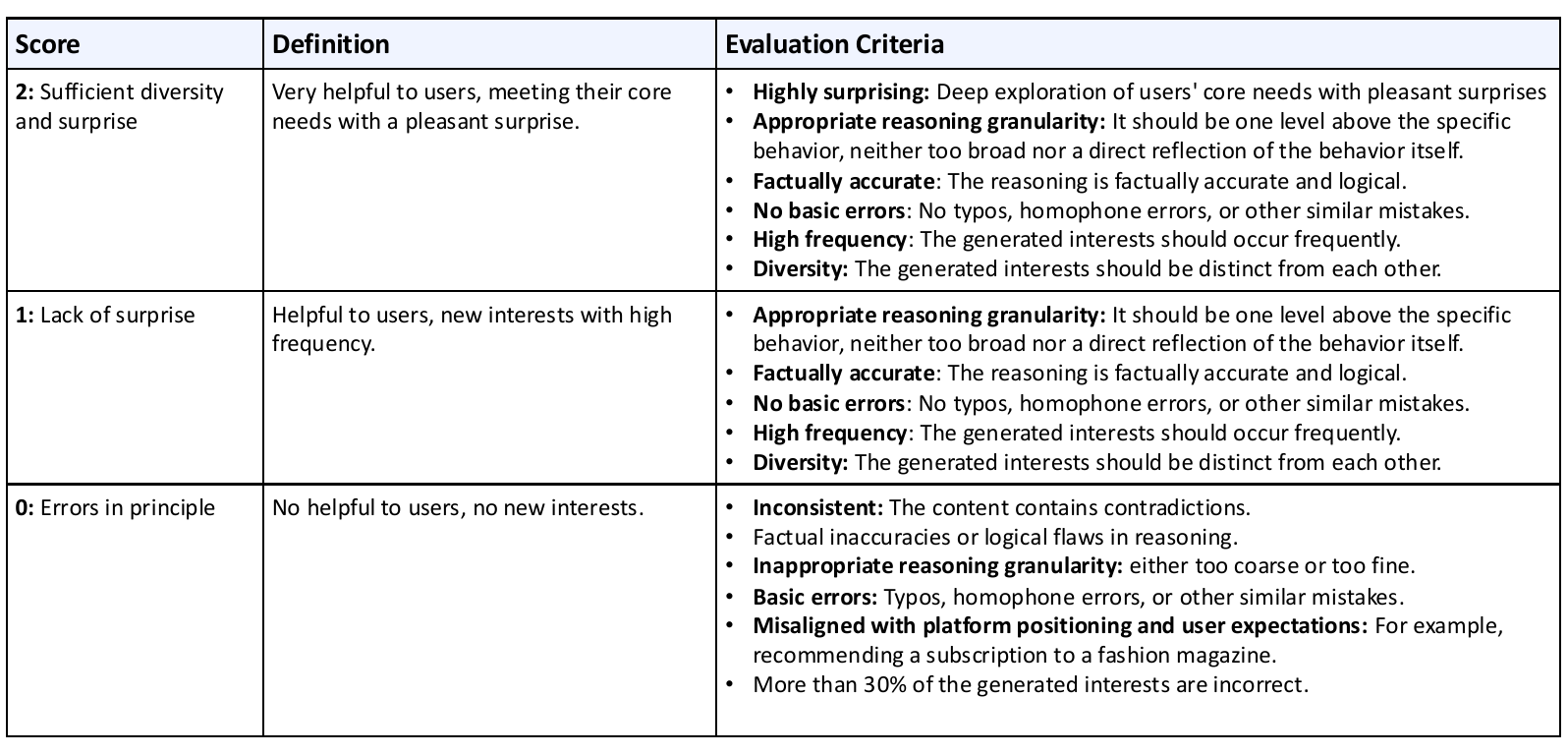}
\caption{Offline evaluation criteria of user potential interests.}
\label{evaluation}
\end{figure*}
\subsubsection{Input Layer}
The input features of the user tower include: user static profiles, such as age and gender, and user historical interaction sequences, such as categories, brands, and tags. These features are represented as $f^u$ through ID embedding. User interests are obtained through a text encoder to generate the embedding $z$. During the training phase, positive samples of user interests are the items that users have clicked on, while negative samples are other items sampled within the batch. In the inference phase, user interests are generated through two-hop reasoning to identify latent novel interests. The text encoder can be selected from models such as CLIP\cite{refclip}, BERT\cite{ref105}, USE\cite{ref106}, or BGE\cite{ref108}. In our experiments, we chose CLIP as the encoder.

The input features of the item tower include: static features of the items, such as category systems, brands, and tags, which are represented as $f^v$ through ID embedding; and historical interests represented by items that have been clicked on, using CLIP to generate the embedding $z$.

\subsubsection{User tower \& item tower}  
\leavevmode
\newline
\textbf{User Tower: } 
Concatenate the user features $f^u$ and historical interests $z$, and then pass them through two fully connected layers to obtain:
\begin{equation}
    U(z, f^u) = \text{FC}(\text{FC}(\text{concat}(f^u, z))).
\end{equation}
  
\textbf{Item Tower: } 
Concatenate the item features $f^v$ and historical interests $z$, and then pass them through two fully connected layers to obtain:
\begin{equation}
    V(z, f^v) = \text{FC}(\text{FC}(\text{concat}(f^v, z))).
\end{equation}

\begin{table*}[hbtp]
  \caption{Online improvement over non-serendipity baseline.}
  \label{tab_improve}
  \begin{tabular}{rcccccccc}
    \toprule
    Method | & AVDU & UVCTR & ACR & ER &  ACC-1 & ACC-3 & ENR & CNR\\
    \midrule
    Ours \quad| & $+0.15\%$ & $+0.07\%$ & $+0.15\%$ & $+0.30\%$ & $+0.21\%$ & $+0.25\%$ & $+4.62\%$ & $+4.85\%$ \\
    \bottomrule
  \end{tabular}
\end{table*}

\subsubsection{Training Objective}
To train the user click samples using the dual-tower model, we aim to ensure that for the same user, the interest representations obtained from different $z$ inputs into user tower are highly distinguishable: the user interest representation $U(z, f^u)$ under interest $z$ should be more correlated with the item representation under the same interest $z$, and this correlation should be stronger than that between the user interest representation $U(z', f^u)$ under other interests $z'$ and the item representation under interest $z$. This approach ensures that when the user's potential interest is fed into user tower, the representation obtained reflects the user's novel interests without being confused with existing interests.

Therefore, we introduce contrastive learning loss:
\begin{equation}
    \mathcal{L}_{\text{CL}} = \frac{e^{\langle U(z, f^u), V(z, f^v) \rangle}}{e^{\langle U(z, f^u), V(z, f^v) \rangle} + \sum_{z' \neq z, z' \in \mathbb{Z}} e^{\langle U(z', f^u), V(z, f^v) \rangle}}.
\end{equation}

Taking all the above considerations into account, we employ a multi-objective joint training approach with a multi-task loss function composed of the contrastive learning loss and binary cross-entropy loss:
\begin{equation}
\mathcal{L} = \mathcal{L}_{\text{BCE}} + \lambda_1 \mathcal{L}_{\text{CL}} + \lambda_2 \| \Theta \|^2,
\end{equation}
where $\Theta$ is the set of model parameters, and $\lambda_1$ and $\lambda_2$ are hyperparameters.

The binary cross-entropy loss is used to model the user's click preferences for historical items, and its formula is:
\begin{equation}
    \mathcal{L}_{\text{BCE}} = - \left[ y \log(\hat{y}) + (1 - y) \log(1 - \hat{y}) \right],
\end{equation}
where \( \hat{y} \) is the predicted click probability for item.

\subsubsection{Inference}
During the inference phase, we first feed the user's potential new interest $z_k^u$ (where $1 \leq k \leq n$ and $n$ denotes the total number of potential new interests for user $u$) along with user features into the user tower. This process generates the user's new interest representation vector $\hat{u}_k$, as formulated in Eq. (\ref{equu}). Subsequently, we utilize $\hat{u}_k$ to perform embedding retrieval and obtain a set of items, which serves as the retrieval result for the potential interest $z_k^u$. The retrieval results from all potential new interests are then aggregated and combined with outputs from other retrieval channels. This consolidated collection is ultimately delivered to subsequent recommendation pipelines for further processing.
\begin{equation}
    \hat{u}_k=U(f^v,z_k^u).
    \label{equu}
\end{equation}

\section{Experiment}

\subsection{Set Up}
We conducted experiments on the Dewu app, a trendy e-commerce platform with tens of millions of users. We randomly selected 10\% of Dewu users to conduct an A/B test, with the goal of generating users' potential interests based on their historical queries and static profiles, and recommending serendipitous items to them. We chose Dewu's existing community recommendation retrieval system as the baseline and used CLIP as the interest text encoder, adding a new retrieval channel for serendipity recommendation on this basis.

\subsection{Metrics}
We employ eight metrics to evaluate online performance: average duration per user (ADU), unique visitor click-through rate (UVCTR), average consumption rate (ACR), unique visitor interaction penetration (ER), average first-level category clicks per user (ACC-1), average third-level category clicks per user (ACC-3), first-level category novelty exposure rate (ENR), and first-level category novelty click rate (CNR). Among these, ACC-1 and ACC-3 serve as indicators for assessing diversity. We define first-level category novelty as follows: an item's exposure or click is deemed to possess first-level category novelty when its first-level category is not included in the set of first-level category from the user's most recent 200 clicks. By calculating the proportion of first-level category novelty exposures out of all exposures and the proportion of first-level category novelty clicks out of all clicks, we assess the novelty performance of the recommendation system.

\subsection{Offline Results}
We used 30,000 labeled samples generated by DeepSeek-R1\cite{deepseek} to fine-tune the QWQ-32B model via SFT, resulting in the InterestGPT model. Applying the offline evaluation criteria in Figure \ref{evaluation}, we assessed InterestGPT on a test set of 1,000 samples. The evaluation involved a random sample of 1,000 users, yielding the following distribution of scores: 1\% received a score of 0, 3\% received a score of 1, and 96\% received a score of 2.

\subsection{Online A/B Test}
To evaluate the online performance of our method, we randomly selected 10\% of Dewu users for A/B testing. Building on the baseline, we introduced an additional retrieval channel for novelty recommendation. Within this serendipitous retrieval channel, we expanded users' potential interests based on their search behaviors over the past 30 days, selecting up to 16 potential interests per user and retrieving 40 corresponding items for each interest. The retrieval results from this channel were then merged with those from other channels to obtain the final retrieval outcomes.

As shown in Table \ref{tab_improve}, compared to the baseline, our method significantly enhances the diversity and novelty of recommendation results. Our method achieves a 0.15\% improvement in AVDU. UVCTR, ACR, and ER have increased by 0.07\%, 0.15\%, and 0.3\%, respectively. In terms of diversity, ACC-1 and ACC-3 have seen improvements of 0.21\% and 0.23\%, respectively. For novelty, ENR and CNR have achieved significant increases of 4.62\% and 4.85\%, respectively.

The serendipity retrivel channel consistently improves the diversity and novelty of recommended content. The exposure novelty rate for the control group was 14.24\%, while the serendipitous retrieval channel in the experimental group achieved a retrieval novelty rate of 26.53\%, compared to 16.17\% for other channels. This indicates that once the serendipity  retrieval introduces new signals, prompting users to engage in new interactions and generating training data associated with novel interests, other retrieval channels can swiftly capture these new interest signals. Consequently, this disrupts the feedback loop phenomenon and breaks the filter bubble phenomenon.

\section{Conclusion}
This work addresses the filter bubble issue in recommendation systems by proposing a method that leverages large models to construct dynamic user knowledge graphs and performs two-hop reasoning. It consists of two stages: two-hop reasoning, which dynamically constructs user knowledge graphs based on user static profiles and historical behaviors using large language models and conducts two-hop reasoning on the constructed graphs; and near-line adaptation, which facilitates efficient industrial deployment. We propose a u2i retrieval model that also incorporates i2i retrieval capabilities, the retrieved items not only exhibit strong relevance to users' newly emerged interests but also retain the high conversion rate of traditional u2i retrieval. Online experiments demonstrate that the method increased the exposure novelty rate by 4.62\%, the click novelty rate by 4.85\%, the average view duration per person by 0.15\%, unique visitor click-through rate by 0.07\%, and the unique visitor interaction penetration by 0.30\%. Deployed in the "You May Also Like" section of Dewu app, the system showcases the potential of large language models to break the filter bubble and enhance user satisfaction in recommendation systems.

\section{Future Work}
Currently, we have only utilized users' search behaviors within the Dewu app, which are considered sparse actions within the app. It remains to be seen whether incorporating additional behavioral data, such as clicks, can still enable large language models to accurately capture users' potential interests and whether there exists a scaling law for the volume of behavioral data. Beyond using users' new interests in the retrieval stage, we can also integrate users' new interests into the ranking stages, including coarse ranking, fine ranking, and re-ranking, to enhance the scoring accuracy of items under new interests. Additionally, feedback data from the recommendation context can be leveraged to calibrate the multiple interests generated by large models, for instance, to prevent excessive divergence of interests produced by large models and improve the relevance of interest generation.

\bibliographystyle{ACM-Reference-Format}
\bibliography{references}


\end{document}